\documentclass{PoS}

\newcommand{\be}{\begin{equation}}
\newcommand{\bea}{\begin{eqnarray}}
\newcommand{\ee}{\end{equation}}
\newcommand{\eea}{\end{eqnarray}}
\newcommand{\sla}{\slash \hspace{-0.22cm}}

\newcommand{\slak}{\slash \hspace{-0.19cm}}

\def\1eq#1{Eq.~(\ref{#1})}
\def\2eqs#1#2{Eqs.~(\ref{#1}) and~(\ref{#2})}
\def\3eqs#1#2#3{Eqs.~(\ref{#1}), (\ref{#2}) and~(\ref{#3})}
\def\4eqs#1#2#3#4{Eqs.~(\ref{#1}), (\ref{#2}), (\ref{#3}) and~(\ref{#4})}
\def\noeq#1{(\ref{#1})}

\def\fig#1{Fig.~\ref{#1}}

\def\tsigma{\widetilde{\sigma}}

\def\n#1{({\it #1}\hspace{0.02cm})}

\def\G{\Gamma}

\title{Nonperturbative results on the quark-gluon vertex}

\ShortTitle{Nonperturbative results on the quark-gluon vertex}

\author{\speaker{A.~C.~Aguilar}\\
        Universidade Estadual de Campinas - UNICAMP \\
Instituto de F\'{i}sica Gleb Wataghin, 13083-859 - Campinas, SP, Brazil\\ 
        E-mail: \email{aguilar@ifi.unicamp.br}}

\author{{D.~Binosi} \\
European Centre for Theoretical Studies in Nuclear
Physics and Related Areas (ECT*) \\ and Fondazione Bruno Kessler, \\
Villa Tambosi, Strada delle Tabarelle 286, 
I-38123 Villazzano (TN)  Italy \\
E-mail: \email{binosi@ectstar.eu}}

\author{{J.~C. Cardona} \\
Universidade Estadual de Campinas - UNICAMP \\
Instituto de F\'{i}sica Gleb Wataghin, 13083-859 - Campinas, SP, Brazil\\ 
E-mail: \email{jeinerc@ifi.unicamp.br}}

\author{{J. Papavassiliou} \\
Department of Theoretical Physics and IFIC, 
University of Valencia and CSIC, \\
E-46100, Valencia, Spain \\
E-mail: \email{Joannis.Papavassiliou@uv.es}}

\abstract{We present analytical and numerical results for the Dirac  
form factor of the  quark-gluon vertex in the  quark
symmetric limit, where the incoming and outgoing quark momenta have the
same magnitude but opposite sign. To accomplish this, we compute  the relevant components
of the  quark-ghost
scattering kernel at the one-loop dressed approximation, using as 
basic ingredients the full quark propagator,
obtained as a solution of the quark gap equation, and the 
gluon propagator and ghost dressing function, obtained from large-volume lattice simulations.}

\FullConference{Xth Quark Confinement and the Hadron Spectrum,\\
		October 8-12, 2012\\
		TUM Campus Garching, Munich, Germany}

\begin{document}

\section{Introduction}

In the last few years, considerable progress has been made in our understanding 
of the infrared (IR) behavior of the fundamental 
Green's functions of QCD, such as gluon, ghost, and quark 
propagators~\cite{Cucchieri:2007md,Bogolubsky:2007ud,Oliveira:2009eh,Aguilar:2008xm,Binosi:2009qm,Fischer:2006ub,Dudal:2008sp,Roberts:1994dr,Bashir:2011dp}, 
as well as some of the basic vertices of the theory~\cite{Fischer:2006ub,Bhagwat:2004kj,Kizilersu:2006et,Skullerud:2004gp},  and their
relation to the confinement and dynamical chiral symmetry breaking (CSB)~\cite{Roberts:1994dr}.
In fact, there is a broad consensus that one of the most important ingredients for the CSB 
is the non-abelian \emph{quark-gluon  vertex}, which controls the way the ghost  sector enters
into  the  gap  equation.  Specifically,  this  vertex  introduces  a
numerically crucial dependence on  the ghost dressing function and 
the \emph{quark-ghost scattering amplitude}~\cite{Aguilar:2010cn}. This latter quantity 
satisfies its own dynamical equation,
which  may be decomposed  into individual  integral equations  for its
various  form  factors.  
Here we will present the first steps towards the determination
of the longitudinal quark-gluon vertex form factors
for a particular kinematic configuration: the 
\emph{quark symmetric limit} where the incoming and outgoing quark momenta
have the same magnitude and opposite signs. 
To do that, we compute numerically the relevant quark-ghost scattering kernel components at the ``one-loop dressed'' approximation, at the same kinematic point, using as ingredients the  nonperturbative lattice results for the gluon propagator and ghost dressing function of Ref.~\cite{Bogolubsky:2007ud},  and the solution of  the quark gap equation obtained in~\cite{Aguilar:2010cn}  for the full quark propagator.

\section{Ingredients and definitions}

\begin{figure}[!ht]
\begin{center}
\includegraphics[scale=0.55]{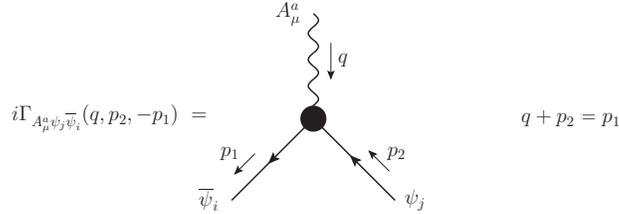}
\caption{\label{quark-vertex}The conventional quark-gluon vertex with the momenta routing used throughout the text.} 
\end{center}
\end{figure}

Consider the conventional quark gluon vertex shown in~\fig{quark-vertex}, and defined according to
\be
i\G_{A^a_\mu\psi_j\overline{\psi}_i}(q,p_2,-p_1)=ig\,t^a_{ij}\,\G_\mu(q,p_2,-p_1);\qquad \G^{(0)}_\mu(q,p_2,-p_1)=\gamma_\mu;\qquad q+p_2=p_1.
\ee
%

In the Batalin-Vilkoviski (BV) formalism, the Slavnov-Taylor identity (STI) satisfied  by this vertex reads (in the kinematic configuration chosen)~\cite{Binosi:2009qm}
\be
q^\mu\G_{A^a_\mu\psi_j\overline{\psi}_i}(q,p_2,-p_1)=
F(q^2)
\left[\G_{\psi_k\overline{\psi}_i}(p_1)\G_{\psi_jc^a\overline{\psi}^*_k}(p_2,q,-p_1)+\G_{\psi^*_k\overline{\psi}_ic^a}(p_2,-p_1,q)\G_{\psi_j\overline{\psi}_k}(p_2)\right].
\label{BV-STI}
\ee
In the formula above $F(q^2)$ denotes the so-called ghost dressing function which is related to the full ghost propagator $D^{ab}(q^2)$ through
\be
iD^{ab}(q^2)=i\delta^{ab}\frac{F(q^2)}{q^2};\qquad F^{(0)}(q^2)=1,
\ee
while $\G_{\psi\overline{\psi}}$ is the inverse of the full fermion propagator $S_{ij}(p)$ obtained by solving the equation 
\be
iS_{ij}(p)\G_{\psi_j\overline{\psi}_k}=\delta_{ik};\qquad iS^{(0)}_{ij}(p)=i\delta_{ij}S^{(0)}(p)=i\delta_{ij}\frac1{p\hspace{-0.18cm}/-m}.
\ee
The standard decomposition for the inverse of the full quark propagator $S^{-1}(p)$ is given by 
\be
iS^{-1}(p) = i[A(p^2)\,\sla{p} - B(p^2)]\,,
\label{qpropAB}
\ee
where $A(p^2)$ and $B(p^2)$ are, respectively, the Dirac vector and scalar
components.

\begin{figure}[!ht]
\begin{center}
\includegraphics[scale=0.55]{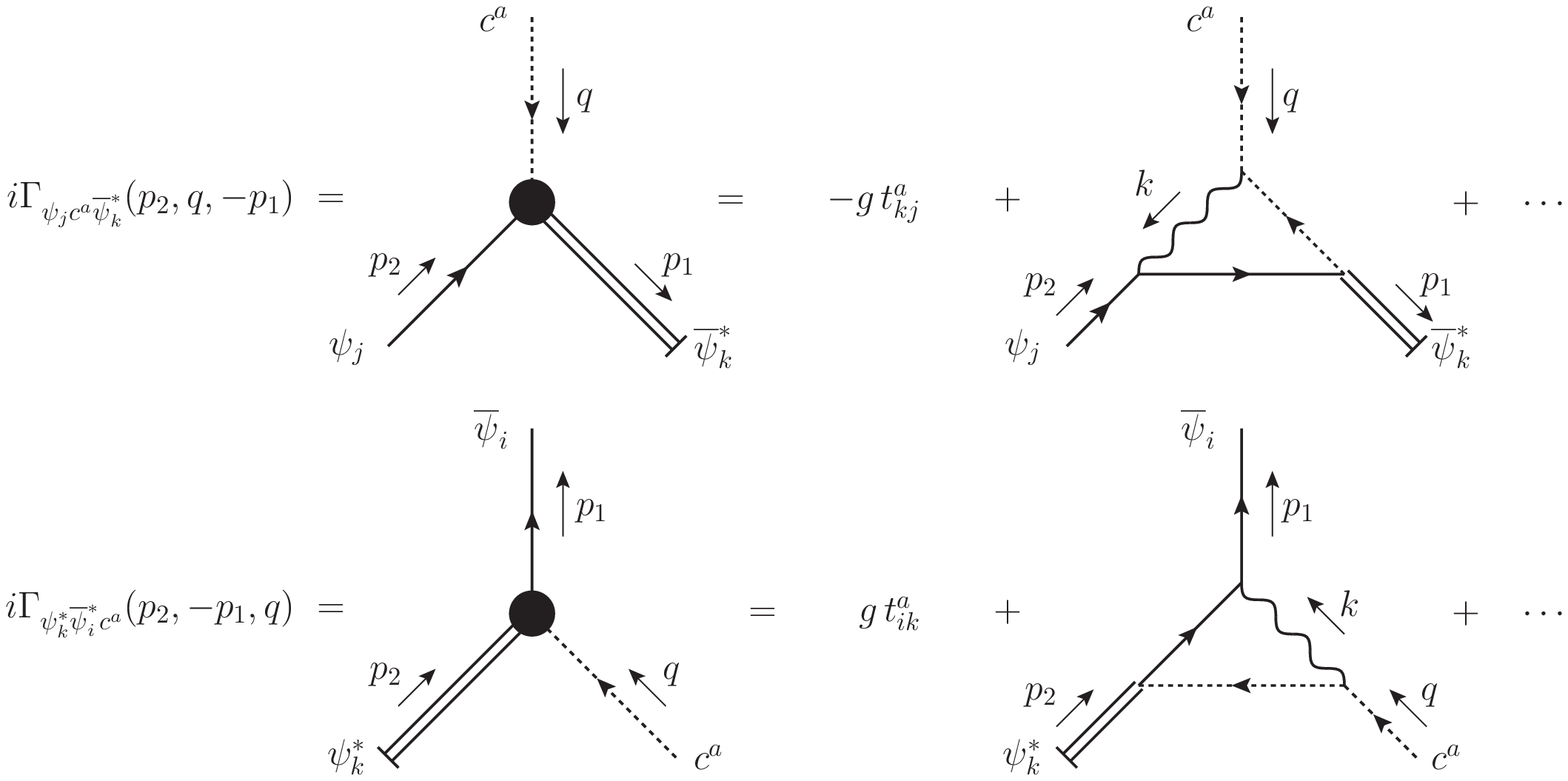}
\caption{\label{aux-functions}The auxiliary functions through which the STI satisfied by the quark-gluon vertex is satisfied. For convenience we show the momenta routing matching the kinematics chosen for the quark-gluon vertex as well as the tree-level and one-loop terms in the corresponding loop expansion.} 
\end{center}
\end{figure}

In addition, $\psi_k^*$ and $\overline{\psi}_k^*$ represent the so-called antifields associated to the spinor fields $\overline{\psi}_k$ and $\psi_k$ respectively; they have ghost charge -1, (mass) dimension 5/2, and obey Bose statistics. 
The Green's functions $\G_{\psi^*_k\overline{\psi}_ic^a}$ and $\G_{\psi_jc^a\overline{\psi}^*_k}$ are shown in~\fig{aux-functions}.

These two functions are not independent, being related by ``conjugation''; indeed, to get one from the other, we need to perform the following operations: \n{i} exchange $-p_1$ with $p_2$: $-p_1\leftrightarrow p_2$; \n{ii} reverse the sign of all external momenta: $q,-p_1,p_2\leftrightarrow-q,p_1,-p_2$; \n{iii} take the hermitian conjugate of the resulting amplitude. 

Then, introducing the function
\bea
H^a_{ij}(q,p_2,-p_1)=g\, t^a_{ij}\, H(q,p_2,-p_1)=-i\G_{\psi_jc^a\overline{\psi}^*_k}(p_2,q,-p_1), \\ \nonumber
\overline{H}^a_{ij}(-q,p_1,-p_2)=g\, t^a_{ij}\, \overline{H}(-q,p_1,-p_2)=i\G_{\psi^*_k\overline{\psi}_ic^a}(p_2,-p_1,q),
\eea
and factoring out a the common color and gauge coupling combination  $g\,t^a_{ij}$, the STI~\noeq{BV-STI} can be rewritten as 
\be
q^\mu\G_{\mu}(q,p_2,-p_1)=F(q^2)\left[S^{-1}(p_1)H(q,p_2,-p_1)-\overline{H}(-q,p_1,-p_2)S^{-1}(p_2)\right],
\label{STI}
\ee
with $\overline{H}$ obtained from $H$ through the set of operations detailed above. 
The $H$ function admits the general
form factor decomposition~\cite{Davydychev:2000rt}
\be
H(q,p_2,-p_1)=X_0 \mathbb{I}+X_1 {p_1\hspace{-0.34cm}/}\hspace{0.1cm}
+X_2p_2\hspace{-0.34cm}/
\hspace{0.1cm}
+X_3\tsigma_{\mu\nu}p^\mu_1p^\nu_2,
\label{HH}
\ee
where the form factors $X_i$ are functions of the momenta, $X_i=X_i(q^2,p_2^2,p_1^2)$ and 
$\tsigma_{\mu\nu}=1/2[\gamma_\mu,\gamma_\nu]$ (notice the $i$ difference with respect to the conventional definition of this quantity). 
One then obtains automatically the expansion
\be
\overline{H}(-q,p_1,-p_2)={\overline{X}}_0 \mathbb{I}+{\overline{X}}_2 {p_1\hspace{-0.34cm}/}\hspace{0.1cm}
+{\overline{X}}_1p_2\hspace{-0.34cm}/
\hspace{0.1cm}
+{\overline{X}}_3\tsigma_{\mu\nu}p^\mu_1p^\nu_2,
\label{HH_bar}
\ee
where $\overline{X_i}={X}_i(q^2,p_1^2,p_2^2)$. 

At tree-level, one clearly has $ X^{(0)}_0= \overline{X}^{(0)}_0=1$, with the remaining form factors vanishing.

The most general Lorentz decomposition for the longitudinal part of the vertex $\Gamma_{\mu}(q,p_2,-p_1)$ 
appearing in Eq.~(\ref{STI})  can be written as~\cite{Davydychev:2000rt}
\be
\Gamma_{\mu}(q,p_2,-p_1) = 
  L_1 \gamma_{\mu}
+ L_2 (\sla{p_1} + \sla{p_2})(p_1+p_2)_{\mu} 
+ L_3 (p_1+p_2)_{\mu} 
+ L_4 \tsigma_{\mu\nu}(p_1+p_2)^{\nu},
\label{Li}
\ee
where $L_i$ are the form factors, whose  dependence on the momenta 
has been suppressed, in order to keep a compact notation, {\it i.e.}, $L_i=L_i(q^2,p_1^2,p_2^2)$. Notice that the tree level expression 
for $\Gamma_{\mu}^{(0)}$  is 
recovered setting $L_1=1$ and $L_2=L_3=L_4=0$; then, 
$\Gamma_{\mu}^{(0)} = \gamma_{\mu}$.

Contracting Eq.~(\ref{Li}) with $q^{\mu}$, we have 
\be
q^{\mu}\Gamma_{\mu} = 
(p_1^2 - p_2^2) L_3 \mathbb{I} 
+[(p_1^2 - p_2^2) L_2 + L_1]\sla{p_1}
+[(p_1^2 - p_2^2) L_2 - L_1] \sla{p_2}
+ 2 L_4 \tsigma_{\mu\nu}p_1^{\mu} p_2^{\nu}.
\label{VLi}
\ee

In addition, substituting into  Eq.~(\ref{STI})  the full quark propagator $S^{-1}(p)$ of
Eq.~(\ref{qpropAB}), and the expressions for $H$  and $\overline{H}$  given by Eqs.~(\ref{HH}) and~(\ref{HH_bar}) respectively, 
we find that the rhs of Eq.~(\ref{VLi}) can be also expressed  in 
terms of the functions $A$, $B$ and $X_i$'s. Then,  it is  relatively straightforward 
to demonstrate that the $L_i$'s may be expressed as~\cite{Aguilar:2010cn}
\bea
L_1 &=& \frac{F(q)}{2} \left\{
A(p_1)[X_0 - (p_1^2+p_1\!\cdot\!p_2)X_3] 
+ A(p_2)[{\overline X}_0 -(p_2^2 + p_1\!\cdot\!p_2){\overline X}_3]\right\} 
\nonumber\\
&+&
\frac{F(q)}{2} \left\{ B(p_1)(X_2-X_1) + B(p_2)({\overline X}_2-{\overline X}_1)\right\};
\nonumber\\
L_2 &=& \frac{F(q)}{2(p_1^2 - p_2^2)} \left\{
A(p_1)[X_0 + (p_1^2 - p_1\!\cdot\!p_2)X_3] 
- A(p_2)[{\overline X}_0 +(p_2^2-p_1\!\cdot\!p_2){\overline X}_3]\right\}
\nonumber\\
&-&
\frac{F(q)}{2(p_1^2 - p_2^2)} \left\{ B(p_1)(X_1+X_2) - B(p_2)({\overline X}_1+{\overline X}_2)\right\};
\nonumber\\
L_3 &=&  \frac{F(q)}{p_1^2 - p_2^2}
\left\{  
A(p_1) \left( p_1^2 X_1 + p_1\!\cdot\!p_2 X_2 \right)
- A(p_2) \left( p_2^2 {\overline X}_1 +p_1\!\cdot\!p_2 {\overline X}_2\right)
- B(p_1)X_0 + B(p_2){\overline X}_0\right\};
\nonumber\\
L_4 &=&\frac{F(q)}{2} \left\{ 
A(p_1) X_2 - A(p_2) {\overline X}_2 - B(p_1) X_3 + B(p_2){\overline X}_3 
\right\}.
\label{expLi}
\eea

It is interesting to notice that setting  in Eq.~(\ref{expLi}) 
$X_0 = {\overline X}_0=1$ and $X_i = {\overline X}_i=0$, 
for $i \geq 1$, and  $F(q) =1$, we obtain  
the  following expressions  

\bea
L_1 = \frac{A(p_1)+A(p_2)}{2}\,,\quad L_2 = \frac{A(p_1)- A(p_2)}{2(p_1^2 - p_2^2)},  \quad L_3 = - \frac{B(p_1)- B(p_2)}{p_1^2 - p_2^2} \,, \qquad  L_4 = 0 \,.
\eea 
which give rise to the so-called Ball-Chiu (BC) vertex~\cite{Ball:1980ay}, widely employed 
in the literature  for studies of CSB~\cite{Roberts:1994dr}. 

\section{The ``one-loop dressed'' approximation for H}

It is clear from Eq.~(\ref{expLi}), that in order to 
determine the  longitudinal form factors $L_i$ , it is necessary to know the nonperturbative behavior of
the form factors $X_i$.

\begin{figure}[!ht]
\begin{center}
\includegraphics[scale=0.55]{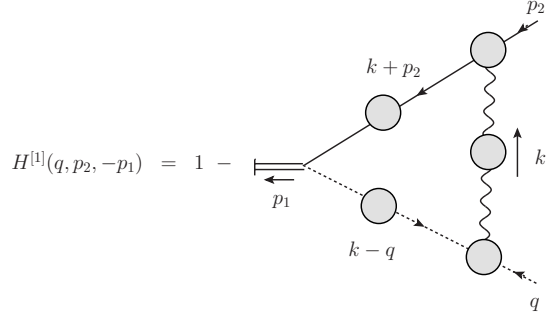}
\caption{\label{skernel}The quark-ghost scattering kernel at ``one-loop dressed'' approximation.} 
\end{center}
\end{figure}

To obtain a nonperturbative estimate for $H$ and its form factors, we
will study the ``one-loop dressed'' contribution represented in the diagram of Fig.~\ref{skernel},
and given by

%
\bea
H^{[1]}(q,p_2,-p_1)=1- \frac{1}{2}\,i\,C_Ag^2\int_k\Delta^{\mu\nu}(k)G_{\nu}(k-q)D(k-q)S(k+p_2)\Gamma_\mu(k,p_2,-k-p_2)\,,
\label{Hexp}
\eea
where $C_A$ is the eigenvalue of the Casimir operator in the adjoint representation, 
$\Delta^{\mu\nu}(q)$ is  the full gluon propagator; in the  Landau gauge  
\bea
i\Delta^{ab}_{\mu\nu}(q)=-i\delta^{ab}\left[g^{\mu\nu}-\frac{q^\mu q^\nu}{q^2}\right]\Delta(q)\,.
\label{propagators}
\eea

For evaluating Eq.~(\ref{Hexp}) further, we will use the following approximations:
(i) the full gluon-ghost vertex will be replaced by its tree-level value, {\it i.e.} $iG^{abc}_{\nu} = -gf^{abc}(k-q)_{\nu}$.
Note that, since  the full $\Delta^{\mu\nu}(k)$ is transverse, 
only the  $q_{\nu}$ part of the gluon-ghost survives since $k_\nu\Delta^{\mu\nu}(k)=0$; (ii) for the vertex $\Gamma_{\mu}$ we will use 
the following Ansatz
\bea
\Gamma_{\mu}(k,p_2,-k-p_2) &=&\frac{F(k)}{2}\left[A(p_2+k)+A(p_2)\right]\gamma_\mu \nonumber \\
&+&\frac{F(k)}{2}\frac{k^\mu}{k^2}\Big[\left[A(p_2+k)-A(p_2)\right]
\left(2\;\sla p_2+\slak k\right)-2\left[B(p_2+k)-B(p_2)\right]\Big]\,.
\label{vertex1}
\eea
 
Notice that the above Ansatz satisfies the STI of Eq.~(\ref{STI}) when $H = 1$. 
Again, due to the transversality of $\Delta_{\mu\nu}(k)$,  the second term on the rhs of Eq.~(\ref{vertex1}), which is proportional to  the longitudinal momentum $k^{\mu}$, does not contribute in the  Eq.~(\ref{Hexp}).

Then,  inserting into Eq.~(\ref{Hexp}) the propagators of Eq.~(\ref{propagators}) and the Ansatz for the quark-gluon vertex given by Eq.~(\ref{vertex1}), it is straightforward to derive the following expression for $H$
\bea
\label{Hgeral}
H^{[1]}(q,p_2,-p_1)=1+i\frac{g^2C_A}{4}\int_k {\mathcal K}(p_2,q,k) f(p_2,q,k)\,,
\label{Hgeneral}
\eea
where 
\bea
{\mathcal K}(p_2,q,k) = \frac{D(k-q)F(k)\Delta(k)[A(p_2+k)+A(p_2)]}{A^2(p_2+k)(p_2+k)^2-B^2(p_2+k)} \,,
\label{kernh}
\eea
while all spinorial structure is included in 
\bea
f(p_2,q,k) = A(p_2+k)\left[ \sla{p_2}\;\sla{q} + \slak{k}\;\sla{q} -k\cdot q  -\sla{p_2}\,\slak{k}\,\frac{q\cdot k}{k^2} \right] 
+ B(p_2+k) \left[\sla{q}-\slak{k}\;\frac{q\cdot k}{k^2} \right]\,. 
\label{kernf}
\eea

\section {Quark symmetric configuration}

The projection of the  form factors $X_i$, appearing in the definition~(\ref{HH}), for arbitrary kinematics boils down to a
complicated system of several equations.  In order to make the problem at hand
technically more tractable, we will only compute it in a specific kinematical limit: 
the \emph{quark symmetric limit} where the quark momenta
have the same magnitude and opposite signs,  {\it i.e.} \mbox{$p_1=-p/2$}, \mbox{$p_2=p/2$} and \mbox{$q=-p$} .

In this kinematical configuration, it is easy to see that only the 
(Dirac) form factor $L_1$ survives, and the vertex of Eq.~(\ref{Li}) simplifies to 
\be
\Gamma_{\mu}(-p,p/2,p/2)=  L_1\gamma_{\mu}\,; \quad\mbox{where}\quad L_1\equiv L_1(p^2,p^2/4,p^2/4) \,.
\label{z2}
\ee
In addition,  the quark-ghost scattering kernel  $H$ and $\overline H$, given by  Eq.~(\ref{HH}) and Eq.~(\ref{HH_bar}),
simplify in this limit, and the terms proportional to $\sla{p}$ become linearly dependent such that   
\bea
H(-p,p/2,p/2) &=& X_0\, \mathbb{I}  
+(X_2-X_1)\, \sla{p}/2  \,, \nonumber \\  
 {\overline H}(p,-p/2,-p/2) &=&
{\overline X}_0 \,\mathbb{I} 
+({\overline X}_1 -{\overline X}_2 )\, \sla{p}/2 \,.
\label{sym_lim}
\eea

Setting \mbox{$p_1=p_2=p/2$} and \mbox{$q=-p$} in Eq.~(\ref{Hgeral}), 
and taking the appropriate traces, it is straightforward to derive the following expressions for
the form factor $X_0$ and the subtraction $X_2- X_1$ (in the Euclidean space)
\bea
X_0(p)&=&1+\frac{g^2C_A}{8}\int_k\frac{D(p+k)F(k)\Delta(k)A_2[A_2+A_1]}{A^2_2(p/2+k)^2+B^2_2}
\left[p^2-\frac{(p\cdot k)^2}{k^2}\right] \,, \nonumber \\
X_2(p)-X_1(p)&=&\frac{g^2C_A}{2p^2}\int_k\frac{D(p+k)F(k)\Delta(k)B_2[A_2+A_1]}{A^2_2(p/2+k)^2+B^2_2}
\left[p^2-\frac{(p\cdot k)^2}{k^2}\right]\,,
\label{fsy}
\eea
where $A_1=A(p/2)$, $A_2=A(p/2+k)$,  $B_1=B(p/2)$, and $B_2=B(p/2+k)$. Due to the fact 
that the momenta $p_1$ and $p_2$ have the same magnitude in this configuration, 
it is possible to show from the definition~(\ref{HH_bar}) that \mbox{$X_0={\overline X}_0$} and 
$X_2-X_1 = {\overline X}_2 -{\overline X}_1$.

Then, it is straightforward to see, from Eq.~({\ref{expLi}}), that in the symmetric quark limit 
$L_1$ simplifies to 
\bea
L_1=F(p)X_0(p)A(p/2) + F(p)[X_2(p)-X_1(p)]B(p/2)\,.
\label{l1_sym}
\eea

As we have seen in Eq.~({\ref{fsy}}), both form factors
$X_0$ and $X_2-X_1$  depend  on the nonperturbative form 
of the four basic Green's functions, namely $\Delta(q)$, $F(q)$, $A(p)$ and $B(p)$.
Therefore, in order to proceed with the numerical analysis, we use 
for $\Delta(q)$ and $F(q)$ the lattice data obtained by ~\cite{Bogolubsky:2007ud}, while
for $A(p)$ and $B(p)$ we use the solution of the quark gap equation
obtained in Ref~\cite{Aguilar:2010cn}.  All these
functions were renormalized at \mbox{$\mu= 4.3$ GeV}, and
in all our calculations we have fixed \mbox{$\alpha(\mu^2)=g^2/4\pi=0.30$}.


\begin{figure}[!ht]
\begin{center}
\begin{minipage}[b]{0.45\textwidth}
\includegraphics[scale=0.47]{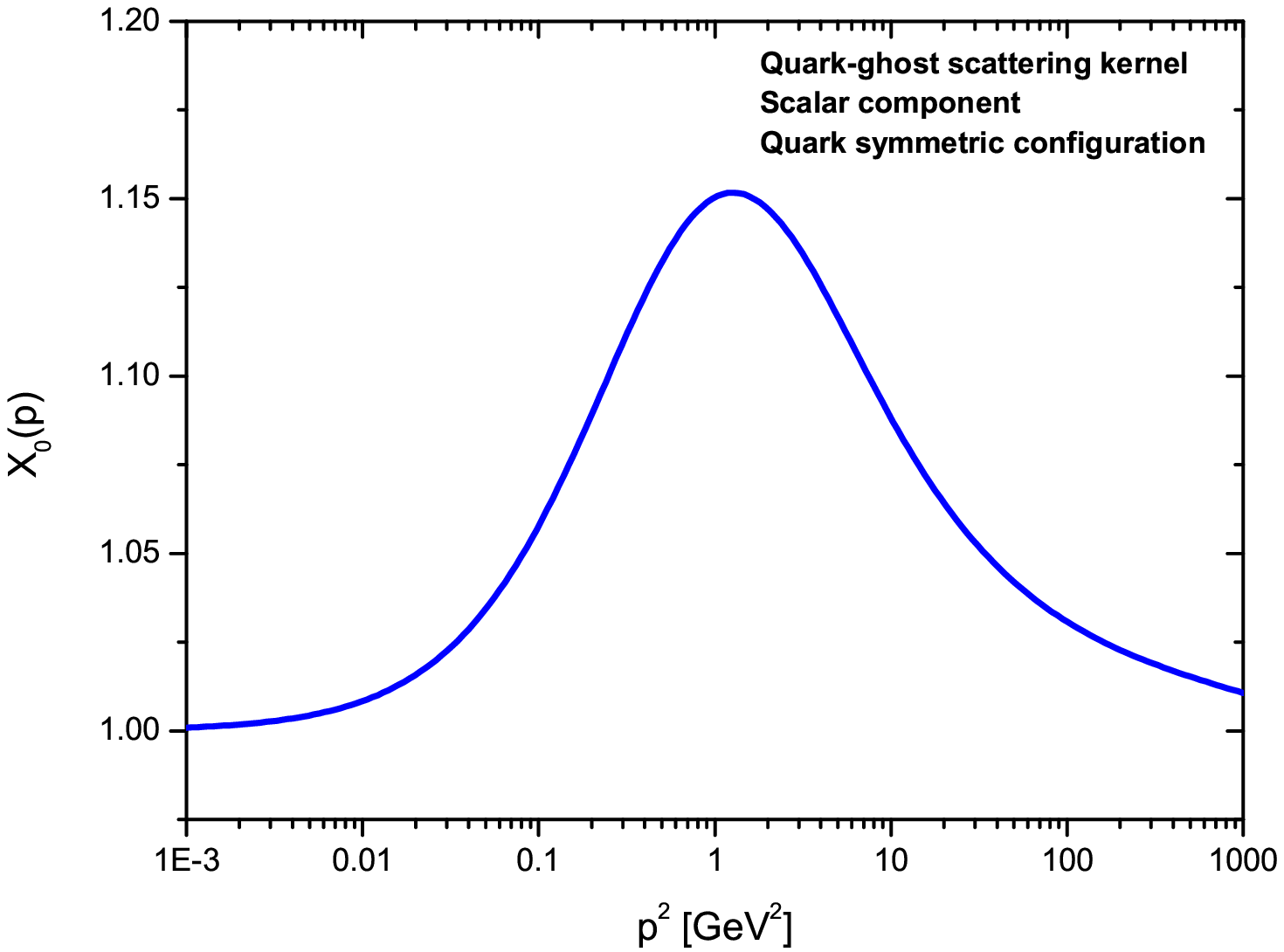}
\end{minipage}
\hspace{0.5cm}
\begin{minipage}[b]{0.50\textwidth}
\includegraphics[scale=0.47]{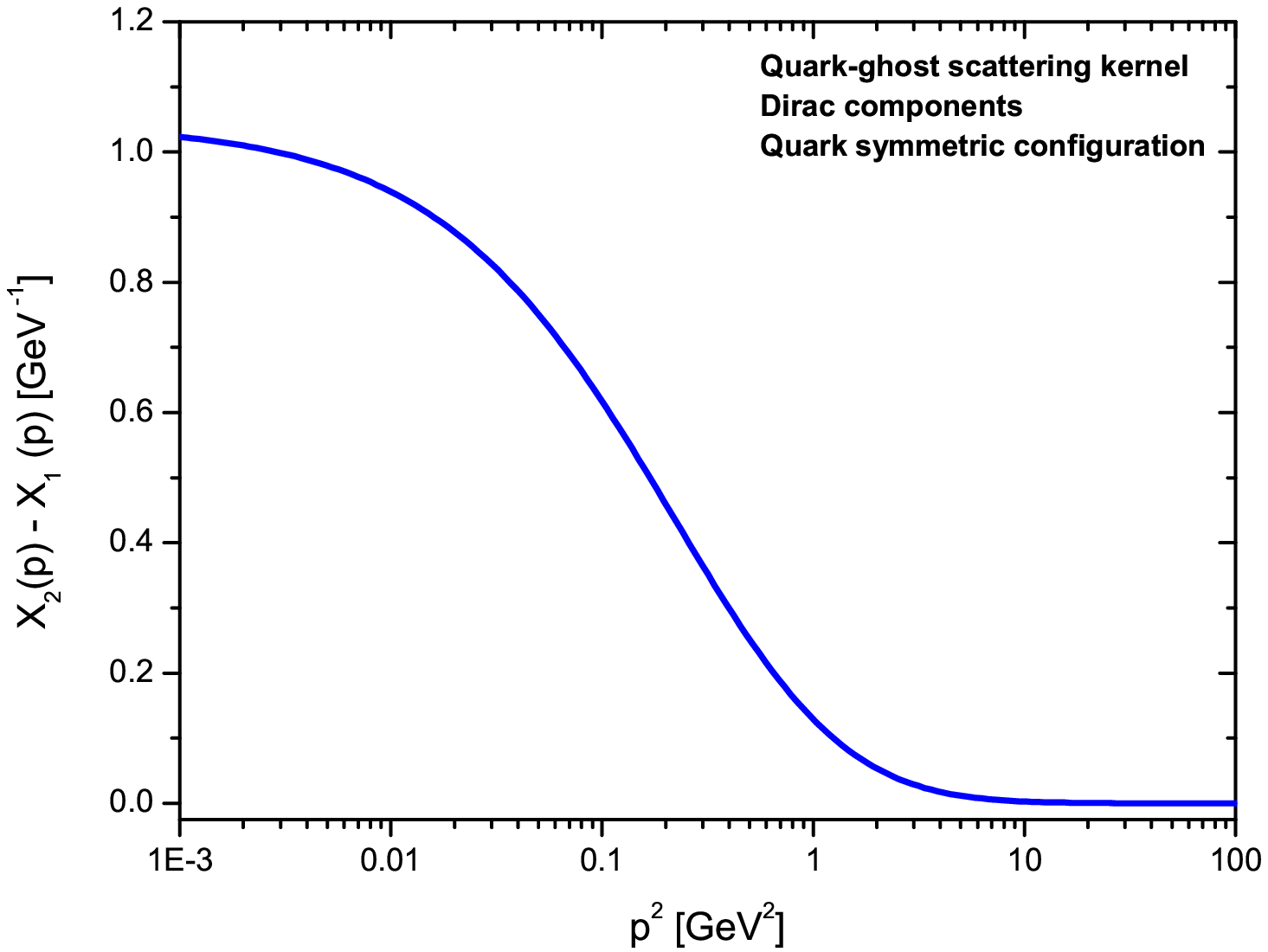}
\end{minipage}
\end{center}
\caption{The scalar form factor $X_0$ (left panel) and 
the combination of the form factors \mbox{$X_2-X_1$}(right panel)  in
the symmetric quark configuration for \mbox{$\alpha(\mu^2)=g^2/4\pi=0.30$}.}
\label{results}
\end{figure}
In Fig.~\ref{results}, we show the numerical results for $X_0(p)$ (left panel), and $X_2(p)-X_1(p)$ (right panel). 
On the left panel of Fig.~\ref{results}, we can see that  $X_0$ shows a maximum  
located in the intermediate momentum region (around $1-2 \,\mbox{GeV}^2$),   
while in the UV and IR regions the curve goes to its perturbative value {\it i.e.} $X_0 \to 1$.

On the right panel of Fig.~\ref{results} we notice that the combination $X_2 -X_1$ saturates
at a finite value in the deep IR region, while in the UV it 
vanishes asymptotically.

With all ingredients available, we are now in position to determine
the behavior of the  Dirac form factor $L_1$ in the symmetric quark configuration.

\begin{figure}[!h]
\begin{center}
\resizebox{0.45\columnwidth}{!}
{\includegraphics{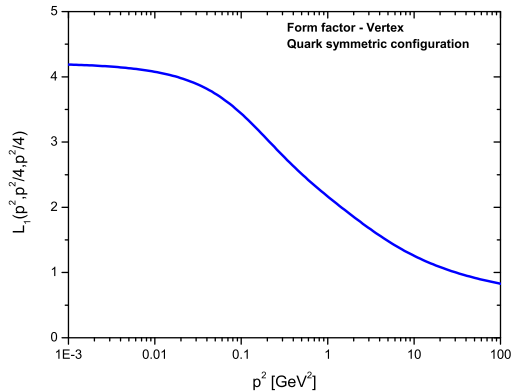}}
\end{center}
\vspace{-0.5cm}
\caption{Numerical result for the vertex form factor $L_1$ 
in the quark symmetric configuration when \mbox{$\alpha(\mu^2)=0.3$}.}
\label{L1_res}
\end{figure}

In Fig.~\ref{L1_res} we show the result for $L_1$ obtained from Eq.~(\ref{l1_sym}). As we can clearly see,
$L1$  develops a sizable plateau in the IR region, and 
as expected, it recovers its perturbative value in the deep UV region.

\section{Conclusions}

We have presented the general methodology for determining the 
longitudinal form factors of the  quark-gluon vertex from the 
fundamental STI that it satisfies. A key ingredient in this analysis is the  
quark-ghost
scattering kernel, $H$ and its conjugate ${\overline H}$, whose field-theoretic origin 
and basic kinematic properties are rather subtle.

The first nonperturbative estimate of the form factors comprising $H$ has been  
computed using the ``one-loop dressed'' approximation of the corresponding  
integral equation, under certain reasonable dynamical assumptions. For the purposes of this 
presentation we have limited our analysis  
to the particular kinematic limit known as ``quark symmetric
configuration'', which gives rise to considerable technical simplifications.
The  Dirac form factor of the quark-gluon vertex, $L_1$, has been obtained 
in this particular kinematic configuration, and its basic features have been studied. 
The methodology presented here may be directly extended to arbitrary kinematic configurations,   
furnishing valuable information on such a fundamental quantity as the  quark-gluon vertex.

\acknowledgments
We would like to thank the organizers of the Xth Quark Confinement and the Hadron Spectrum for the pleasant conference. 
The research of ACA is supported by CNPq under the grants 305850/2009-1, 453118/2010-0, and by FAPESP - grant 2012/15643-1. 
The work of J.~P. is supported by the Spanish MEYC under 
grant FPA2011-23596.

\end{document}